\documentclass[12pt, a4paper]{article}
\pdfoutput=1

\usepackage[height=25cm,width=16cm]{geometry}
\usepackage{amsfonts}
\usepackage{amsmath}
\usepackage{amssymb}
\usepackage{ascmac}
\usepackage{dcolumn}
\usepackage{bm,here}
\usepackage{subfig}
\usepackage{comment}
\usepackage{ifpdf}
\usepackage{slashed}
\usepackage{colortbl}
\usepackage{color}
\usepackage[mathscr]{eucal}
\usepackage[sort&compress, numbers, merge]{natbib}

\ifpdf
  \usepackage{graphicx}     
  \usepackage[bookmarksopen,colorlinks=true,linkcolor=bblue,citecolor=ppink,urlcolor=ppink]{hyperref}
\else     
  \usepackage[dvipdfmx]{graphicx}     
  \usepackage[dvipdfmx,bookmarksopen,colorlinks=true,linkcolor=bblue,citecolor=ppink,urlcolor=ppink]{hyperref}
\fi

\usepackage{multicol}
\definecolor{red}{rgb}{1,0,0}
\definecolor{ppink}{rgb}{0.921545,0.440586,0.687243}
\definecolor{bblue}{rgb}{0.400000,0.400000,1.000000}
\usepackage[charter]{mathdesign}

\begin{document}

\begin{titlepage}

\begin{flushright}
IPMU 16-0024
\end{flushright}

\begin{center}

\vskip 3.0cm
{\large \bf Heavy Fermion Bound States for Diphoton Excess at 750GeV\\
\vspace{0.3cm}
$\sim$ Collider and Cosmological Constraints $\sim$}

\vskip 1.5cm
{\large
Chengcheng Han$^{(a)}$,
Koji Ichikawa$^{(a)}$,
Shigeki Matsumoto$^{(a)}$,
\\[0.5em]
Mihoko M. Nojiri$^{(a, b, c)}$
and
Michihisa Takeuchi$^{(a)}$
}

\vskip 2.0cm
\begin{tabular}{l}
$^{(a)}$ {\em Kavli IPMU (WPI), UTIAS, University of Tokyo, Kashiwa, 277-8583, Japan}
\\[0.3em]
$^{(b)}$ {\em KEK Theory Center, IPNS, KEK, Tsukuba, 305-0801, Japan}
\\[0.3em]
$^{(c)}$ {\em Graduate University of Advanced Studies (Sokendai),Tsukuba, 305-0801, Japan}
\end{tabular}

\vskip 2.0cm
\begin{abstract}
\noindent
A colored heavy particle with sufficiently small width may form non-relativistic bound states when they are produced at the large hadron collider\,(LHC), and they can annihilate into a diphoton final state. The invariant mass of the diphoton would be around twice of the colored particle mass. In this paper, we study if such bound state can be responsible for the 750\,GeV diphoton excess reported by ATLAS and CMS. We found that the best-fit signal cross section is obtained for the SU(2)$_L$ singlet colored fermion $X$ with $Y_X=4/3$. Having such an exotic hypercharge, the particle is expected to decay through some higher dimensional operators, consistent with the small width assumption. The decay of $X$ may involve a stable particle $\chi$, if both $X$ and $\chi$ are odd under some conserved $Z_2$ symmetry. In that case, the particle $X$ suffers from the constraints of jets + missing $E_T$ searches by ATLAS and CMS at 8\,TeV and 13\,TeV. We found that such a scenario still survives if the mass difference between $X$ and $\chi$ is above $\sim$ 30\,GeV for $m_X \sim 375$~GeV. Even assuming pair annihilation of $\chi$ is small, the relic density of $\chi$ is small enough if the mass difference between $X$ and $\chi$ is smaller than $\sim$ 40\,GeV.
\end{abstract}

\end{center}

\end{titlepage}

\section{Introduction}
\label{sec: intro}

The LHC Run II at 13\,TeV has started last year and first results have been obtained. Among observed deviations from standard model\,(SM) predictions, the excess of diphoton events with an invariant mass around 750\,GeV has been reported by both ATLAS and CMS collaborations\,\cite{ATLAS, CMS:2015dxe}. The global significance of the excess is 2.3\,$\sigma$\,(2$\,\sigma$) for ATLAS\,(CMS), while the local significance is 3.6$\,\sigma$\,(2.6$\,\sigma$). The best fit value of the decay width is around 45\,GeV for ATLAS data, while CMS data is more significant in the narrow width approximation. If one interprets the excess as a resonance of an unknown particle, the cross section times the branching ratio to the diphoton channel is required to be around 5\,fb\,\cite{Buttazzo:2015txu, Franceschini:2015kwy, Ellis:2015oso, Low:2015qep, Falkowski:2015swt}.

A possible explanation of the excess by a spin zero resonance (scalar or pseudoscalar) has been extensively studied in recent literatures. On the other hand, existing negative search results at the 8\,TeV LHC constrain the nature of the observed excess. The upper limit of the production cross section at 8\,TeV is scaled to a constraint at 13\,TeV using the ratio of luminosity functions at 8\,TeV and 13\,TeV. The production cross section of the (pseudo)scalar particle should increase by a factor of 2.5 at 13\,TeV if the production though $q\bar{q}$ initial states dominates, while it increases by 4.5 if the gluon gluon fusion dominates. Since the search of the diphoton resonance at 8\,TeV gives a stringent upper limit on the production cross section of 1--2\,fb, the production of the (pseudo)scalar particle through the gluon gluon fusion should be the dominant mechanism of the excess\,\cite{Buttazzo:2015txu, Franceschini:2015kwy, Ellis:2015oso, Low:2015qep, Falkowski:2015swt}. Note that the new (pseudo)scalar particle couples to gluons or photons at loop level, while it is difficult to explain the excess if only SM particles are involved in the loop. Colored vector fermions or scalars should be introduced to explain the observed excess. $O(1)$ couplings between the (pseudo)scalar and the new colored particles are needed to have a sufficient cross section, though these couplings could blow up at the scale not much beyond TeV\,\cite{Hall:2015xds, Zhang:2015uuo, Gu:2015lxj, Bae:2016xni}.

On the other hand, the resonance could arise naturally as a bound state of a new colored particle $X$ when the decay width of the particle is small enough. The colored particles in the bound state can annihilate into gauge bosons, so that they can give a relatively clean diphoton signature. The possibility to observe such a resonance at hadron colliders has been studied extensively for the scalar top case in the past\,\cite{Herrero:1987df, Barger:1988sp, Drees:1993yr, Drees:1993uw, Martin:2009dj, Kats:2009bv, An:2015uwa} and in the context of the 750\,GeV excess\,\cite{Luo:2015yio, Potter:2016psi, NewOne}. However, such a colored particle of the mass $\sim 375$\,GeV ($750/2$\,GeV) is severely constrained by the current LHC data. For example, it is excluded up to 750\,GeV if the decay modes consist of $tZ$, $tH$ and $bW$ for the fermionic top partner case\,\cite{ATLAS:2015dka, Chatrchyan:2013uxa}. One way to evade such current searches is introducing a dark matter particle $\chi$ in the decay chain of $X$. Even for the case, when $m_X - m_\chi$ is large enough, scalar top searches or general SUSY searches exclude $m_X >$ 900\,GeV (700\,GeV) for the fermion (scalar) $X$ case\,\cite{Anandakrishnan:2015yfa, Aad:2014bva, CMS:2015kza, Khachatryan:2015pwa, Aad:2014kra}. We therefore consider a degenerate spectrum with smaller $m_X - m_\chi$ as it is well known that the collider sensitivity becomes weaker. It is also preferable to explain its small decay width to enhance the diphoton signal strength. Note that an extremely small width predicts a long-lived colored particle and again strongly constrained by $R$-hadron searches\,\cite{Aad:2013gva, ATLAS:2014fka}. In this paper, we consider a scenario with a multiplicatively conserved $Z_2$ symmetry, and assume that a new colored SU(2)$_L$ singlet and $Z_2$ odd fermion $X$ with hypercharge $Y_X$ decays into a stable and neutral $Z_2$ odd particle $\chi$ through higher dimensional operators. We show that our scenario can explain the diphoton excess without conflicting with any other 8\,TeV and 13\,TeV data and the cosmological constraint on the thermal relic density of $\chi$.

This paper is organized as follows: In section\,\ref{sec: quarkonium}, we study the production and the decay of the bound state at the LHC. We solve the Schr\"{o}dinger equation taking the effect of $Y_X$ into account, and obtain the wave function of the bound state. The cross section of the diphoton signal turns out to be sensitive to $Y_X$, and found to be consistent with the excess when $Y_X=4/3$. In section\,\ref{sec: bounds}, we consider the case where $X$ decays into a dark matter particle $\chi$ and multiple jets, and study the current collider bound on $X$ by reinterpreting SUSY searches at the 8\,TeV and 13\,TeV LHC. We found that the current 13 TeV data have already set the strongest constraint on $X$ and $\chi$ with $m_X \sim 375$\,GeV, however $m_X - m_\chi \gtrsim 30$\,GeV have not been excluded yet. We will show in section\,\ref{sec: cosmology} such a mass difference is preferable from a cosmological viewpoint. Assuming the self-annihilation of $\chi$ does not alter its thermal relic density significantly, $m_X - m_\chi \lesssim 40$\,GeV is indeed required for $m_X \sim 375$\,GeV. We briefly mention the outlook of our scenario in section\,\ref{sec: outlook}.

\section{Quarkonium Productions and Decays}
\label{sec: quarkonium}

Various types of colored heavy fermions can contribute to the diphoton excess through their non-relativistic bound states. We adopt in this paper the fermion which is odd under the $Z_2$ symmetry, triplet under SU(3)$_c$, singlet under SU(2)$_L$ and has a hypercharge $Y_X$. Essential part of the Lagrangian describing this heavy fermion $X$ is then given by
\begin{eqnarray}
	{\cal L}_X = \bar{X} (i \slashed{D} - m_X)\,X + \cdots,
\end{eqnarray}
where the covariant derivative is defined as $\slashed{D} = D_\mu \gamma^\mu$ and $D_\mu = \partial_\mu + i g_s\,(\lambda^a/2)\,G_\mu^a + i g' Y_X B_\mu$ with $G_\mu^a$ and $B_\mu$ being the gluon and U(1)$_Y$ gauge boson fields and their corresponding gauge couplings are $g_s$ and $g'$, respectively. The mass of $X$ is denoted by $m_X$. The fermion $X$ have sufficiently short lifetime so that one can avoid stringent constraints from long-lived colored particle searches at the LHC\,\cite{Aad:2013gva, ATLAS:2014fka} and to be consistent with cosmology. The above Lagrangian should thus involve some interactions inducing such a decay, and will be discussed in the next section, because those are not relevant to the discussion here.

When the heavy fermion $X$ is pair produced near the threshold energy, the pair forms a bound-state with a significantly enhanced production cross section. The diphoton process at the LHC, $pp \to X\bar{X} \to \gamma\gamma$, is induced dominantly through the production of the lowest $^1S_0$\,($J^{PC} = 0^{-+}$) state, which is denoted by $S_0$ in our paper. The $\gamma\gamma$ production cross section through the bound state $S_0$ is in the lowest order calculation computed as
\begin{eqnarray}
	\sigma(pp \to S_0 \to \gamma\gamma) =
	\frac{K}{s\,m_{S_0}} \frac{\Gamma_{\gamma\gamma}\,\Gamma_{gg}}{\Gamma_{\rm tot}}
	\left[ \frac{\pi}{8} \int dx_1 dx_2\,\delta(x_1 x_2 - m_{S_0}^2/s)\,f_g(x_1)\,f_g(x_2) \right],
	\label{eq: cross section}
\end{eqnarray}
where $m_{S_0}$ denotes the mass of the bound-state, $s$ is the center-of-mass energy squared and $f_g(x)$ is the gluon parton distribution function (PDF) inside a proton. We adopt the PDF of MSTW2008NLO\,\cite{Martin:2009iq}, where the parenthesis of the right-hand side takes a value of about 2137 at $\sqrt{s} = $13\,TeV when $m_{S_0} =$ 750\,GeV\,\cite{Franceschini:2015kwy}. The so-called $K$-factor, which is introduced to take higher order corrections into account, is denoted by $K$ in the above formula, and is fixed to be two in our analysis.\footnote{There are two contributions to the $K$ factor. First one is from perturbative QCD corrections and it enhances the cross section by $\sim 50\%$\,\cite{Kuhn:1992qw}. The other one is from excited $^1S_0$ bound states and those give another $\sim 50\%$ enhancement\,\cite{Younkin:2009zn}. Contribution from continuum states above the threshold is negligible when $Y_X \lesssim 2$\,\cite{Chway:2015lzg}.} Total decay width of $S_0$ and its partial decay widths into photons and gluons are denoted by $\Gamma_{\rm tot}$, $\Gamma_{\gamma\gamma}$ and $\Gamma_{gg}$, respectively. The widths are given by the wave function of the bound state at the origin, $\psi_0(0)$, as follows\,\cite{Hagiwara:1990sq}:
\begin{eqnarray}
	&& \Gamma_{\rm tot} = \Gamma_{\gamma\gamma}/c_W^4 + \Gamma_{gg} + 2\Gamma_X,
	\label{eq: tot width} \\
	&& \Gamma_{\gamma \gamma} = 48 \pi Y_X^4 \alpha^2\,|\psi_0(0)|^2/m_{S_0}^2, \rule{0ex}{3ex}\\
	&& \Gamma_{gg} = 32 \pi \alpha_s^2\,|\psi_0(0)|^2/(3 m_{S_0}^2), \rule{0ex}{3ex}
\end{eqnarray}
where $c_W \equiv \cos \theta_W$ is the Weinberg angle, $\alpha_s = g_s^2/(4\pi)$, and $\alpha$ is the fine structure constant, respectively. The width $\Gamma_X$ in Eq.\,(\ref{eq: tot width}) is the total decay width of the heavy fermion $X$ and it is assumed to be smaller enough than other two terms $\Gamma_{\gamma\gamma}/c_W^4$ and $\Gamma_{gg}$, which corresponds to $\Gamma_X \lesssim {\cal O}(1)$\,MeV. This assumption will be discussed in the next section. Since $\Gamma_{\gamma\gamma} \ll \Gamma_{gg}$ when $Y_X \sim {\cal O}(1)$, the signal cross section $\sigma(pp \to S_0 \to \gamma\gamma)$ is proportional to $Y_X^4$.

The wave function and the mass of the bound state, $\psi_0(0)$ and $m_{S_0}$, must be determined to evaluate the cross section. They are obtained by solving the Schr\"odinger equation:
\begin{eqnarray}
	\left[-\frac{ \nabla^2_{\bm{r}} }{m_X} + V(\bm{r}) - E_0 \right] \psi_0(\bm{r}) = 0,
	\label{eq: Schrodinger eq}
\end{eqnarray}
where $E_0$ is the energy eigenvalue of the state, and thus the mass of the bound state is given by $m_{S_0} = 2 m_X + E_0$. Here, the wave function is normalization to be $\int d^3 \bm{r}\,\psi_0^*(\bm{r})\,\psi_0(\bm{r}) = 1$.
The potential $V(\bm{r})$ is composed of two different long-range interactions; one is from the strong force and the other is from the Coulomb force, so that it is expressed as
\begin{eqnarray}
	V(\bm{r}) = -Y_X^2\,\frac{\alpha}{|\bm{r}|} + V_{\rm QCD}(|\bm{r}|).
\end{eqnarray}
The explicit form of the QCD potential $V_{\rm QCD}(|\bm{r}|)$ is found in Ref.\,\cite{Hagiwara:1990sq}, which includes the scale dependence of $\alpha_s$ at a short distance as well as the long-range (non-perturbative) QCD effect.\footnote{Honestly speaking, the long-range (non-perturbative) QCD effect is negligible in our study, for the typical Bohr radius of the bound-state $S_0$ is sufficiently small thanks to the mass scale of the heavy fermion $X$.} It is worth emphasizing that the Coulomb force contribution gives a sizable correction to the potential. It enhances the wave function $|\psi_0(0)|$ by 10--30\% when $Y_X \gtrsim 1$. For instance, $|\psi_{0}(0)| \simeq$ 88, 90, 94, 99, 105 and 113\,GeV$^{1.5}$ when $Y_X =$ 0, 2/3, 1, 4/3, 5/3 and 2, respectively, with $m_{S_0}$ being 750\,GeV. We discuss it in more details in appendix\,\ref{app: wave function} together with a useful fitting function of $|\psi_0(0)|$ for various values of $Y_X$.

\begin{figure}[t]
	\centering
	\includegraphics[width=0.85\linewidth]{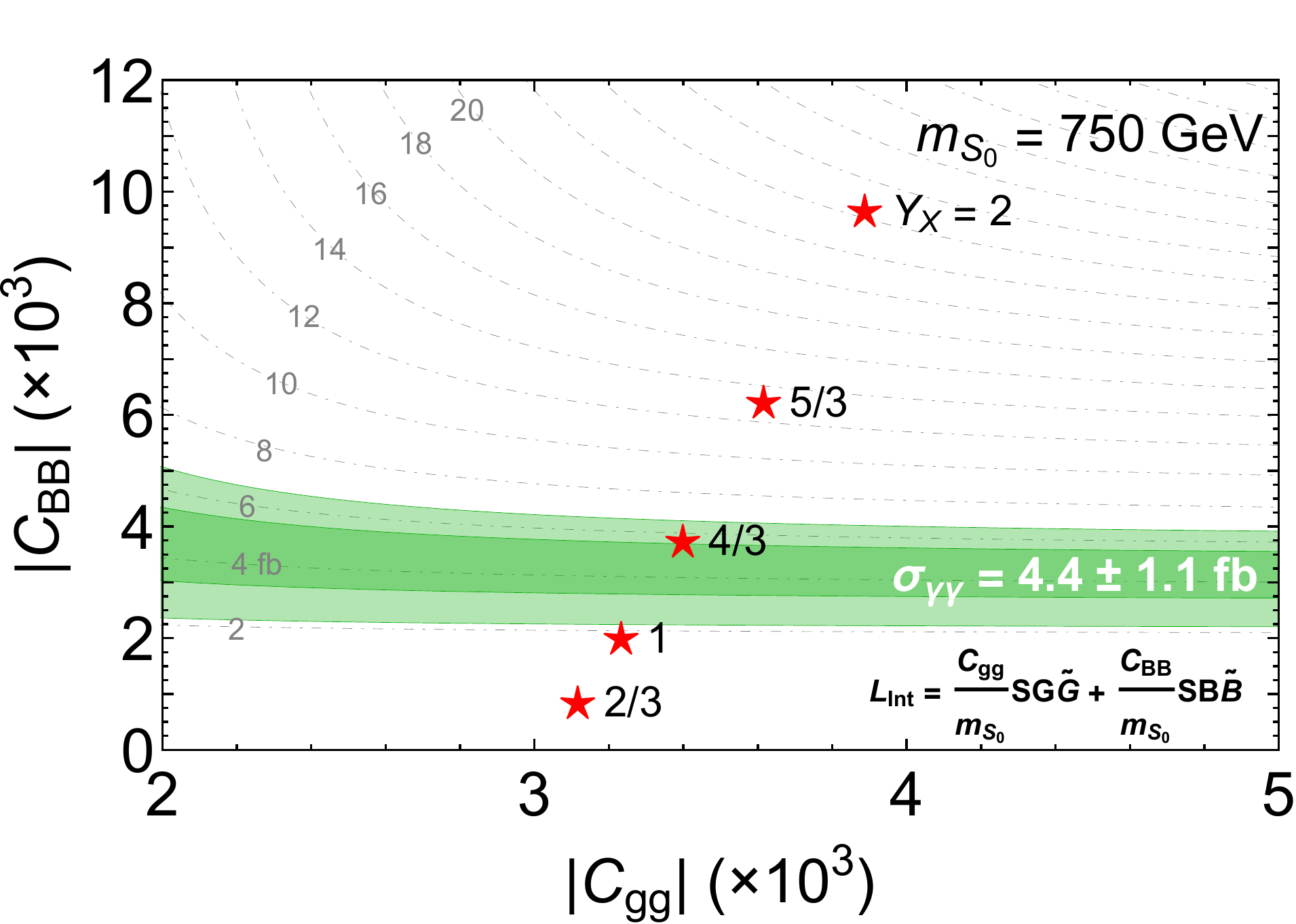}
	\caption{\sl \small Red stars are predictions of our model on the $(|C_{gg}|, |C_{BB}|)$-plane with $Y_X$ being 2/3, 1, 4/3, 5/3 and 2, respectively. Contours of the diphoton cross section as a function of $|C_{gg}|$ and $|C_{BB}|$ are also shown by gray-dashed lines. Darker\,(lighter) green-shaded region corresponds to the cross section experimentally favored by the diphoton excess at 1$\sigma$\,(2$\sigma$) level\,\cite{Buttazzo:2015txu}.}
	\label{fig: CggCBB}
\end{figure}

It is instructive to express the above result in terms of the effective lagrangian ${\cal L}_{\rm eff}$, for ${\cal L}_{\rm eff}$ is frequently used to discuss the diphoton excess from phenomenological viewpoints. Since the bound state $S_0$ is a pseudo-scalar particle composed of a pair of SU(2)$_L$ singlet fermions, ${\cal L}_{\rm eff}$ should involve following dimension-five interactions at leading order:
\begin{eqnarray}
	{\cal L}_{\rm eff} \supset
	\frac{C_{BB}}{m_{S_0}} S_0\,B_{\mu\nu}\,\tilde{B}^{\mu\nu} +
	\frac{C_{gg}}{m_{S_0}} S_0\,G^{a\mu\nu}\,\tilde{G}^a_{\mu\nu},
	\label{eq: effective lagrangian}
\end{eqnarray}
where $B_{\mu\nu}\,(G^a_{\mu\nu})$ and $\tilde{B}_{\mu\nu}\,(\tilde{G}^a_{\mu\nu})$ are the field strength tensor of the U(1)$_Y$\,(SU(3)$_c$) gauge boson and its dual. Matching the effective lagrangian\,(\ref{eq: effective lagrangian}) with the diphoton cross section\,(\ref{eq: cross section}), the absolute values of the coefficients, $|C_{BB}|$ and $|C_{gg}|$, turn out to be $(4 \pi \Gamma_{\gamma\gamma}/c_W^4 m_{S_0})^{1/2}$ and $(\pi \Gamma_{gg}/2 m_{S_0})^{1/2}$, respectively. This fact means that the diphoton signal strength is uniquely determined in our model when the hypercharge $Y_X$ is fixed. This result is shown in Fig.\,\ref{fig: CggCBB}, where the predictions of our model are depicted by red stars on the $(|C_{gg}|, |C_{BB}|$)-plane. For comparison, we also show contours of the diphoton cross section by grey-dashed lines as a function of $|C_{gg}|$ and $|C_{BB}|$. The region painted by a darker (lighter) green color corresponds to the one favored by the diphoton excess at 1$\sigma$\,(2$\sigma$) level\,\cite{Buttazzo:2015txu}. It is worth notifying that the increase of $|C_{gg}|$ with respect to $Y_X$ is from the hypercharge dependence of the wave function, $|\psi_0(0)|$. It can be seen that the heavy fermion with $Y_X = 4/3$ explains the diphoton excess very well, so that will use it as a canonical model in following discussions.

\subsection{Other bound state signals}

When the bound state $S_0$ is produced, it can decay into other channels, which have also been searched for at the 8\,TeV LHC. Since $S_0$ is composed of SU(2)$_L$ singlet fermions, it does not decay into $W^+W^-$ but into $Z\gamma$, $ZZ$ due to the electroweak symmetry breaking. Production cross sections of the channels at 8\,TeV, $\sigma(pp \to S_0 \to Z\gamma, ZZ)$, and experimental limits on the cross sections obtained from 8\,TeV data are shown in Table\,\ref{tab: other decays} with $m_X$ and $Y_X$ being 750\,GeV and 4/3, respectively.\footnote{All the production cross sections in the table have been computed at leading order, namely with the $K$-factor being one, because the cross sections are already much below the experimental limits at the 8\,TeV LHC.} Experimental limits on both of the channels are still weak, though the $Z\gamma$ channel will be important to test the model at the 13\,TeV LHC.

At the threshold energy, the $pp$ collision also produces a bound state which is color neutral but has a spin one with quantum numbers, $^3S_1$\,($J^{PC} = 0^{-\,-}$), which is denoted by $S_1$ in this paper.\footnote{There are no color-octet bound states, because the strong SU(3)$_c$ interaction acts as a repulsive force.} The bound state $S_1$ degenerates with $S_0$ in mass, and is produced dominantly through the s-channel diagram of $\gamma/Z$ from quark-antiquark collisions at leading order. The bound state $S_1$ decays into various fermion pairs. Among those, the decay into a lepton pair ($\ell^+ \ell^- = e^+ e^- + \mu^+ \mu^-$ or $\tau^+ \tau^-$) gives the most sensitive limit, while the next one is the $t\bar{t}$ channel. Their production cross sections and experimental limits from 8\,TeV data are shown in Table\,\ref{tab: other decays}. Since the cross sections are smaller than those of $S_0$, 8\,TeV limits are weak, though the lepton channel would serve another test of the model in future.

\begin{table}[t]
	\centering
	\begin{tabular}{l|rrr}
	\quad\quad @8TeV & Prediction & Limit & Reference \\
	\hline
	$\quad\quad\,S_0 \to Z\gamma$ & 0.74\,fb & 4.0\,fb & \cite{Aad:2014fha} \\
	$\quad\quad\quad\,\,\to ZZ$ & 0.11\,fb & 12\,fb & \cite{Aad:2015kna} \\
	$\quad\quad\,S_1 \to \ell^+ \ell^-$ & 0.13\,fb & 1.2\,fb & \cite{Aad:2014cka} \\
	$\quad\quad\quad\,\,\to \tau^+ \tau^-$ & 0.064\,fb & 12\,fb & \cite{Aad:2014vgg} \\
	$\quad\quad\quad\,\,\to t\bar{t}$ & 0.072\,fb & 550\,fb & \cite{Chatrchyan:2013lca} \\
	$\quad\quad\quad\,\,\to b\bar{b}$ & 0.021\,fb & 1\,pb & \cite{Khachatryan:2015tra} \\
	$S_0 + S_1 \to jj$ & 7\,fb & 2.5\,pb & \cite{CMS:2015neg, Aad:2014aqa} \\
	\hline
	\end{tabular}
	\caption{\sl \small Production cross sections predicted by our model with $m_X$ and $Y_X$ being 750\,GeV and 4/3, respectively, and experimental limits on those from the 8\,TeV LHC data. See text for more details.}
	\label{tab: other decays}
\end{table}

Before closing this section, we also address other signals from $S_0$ and $S_1$. Both of the bound states decay into two jets; a gluon pair from $S_0$, while light quark ($u$, $d$, $s$ and $c$) pairs from $S_1$. Since the total production cross sections of the two jet channel are still much smaller than the experimental limit as shown in Table\,\ref{tab: other decays}, this channel is useless to test the model. The bound state $S_1$ can decay into a bottom quark pair, though it is less significant than the $t\bar{t}$ channel as can be seen in the table. This bound state also decays into $W^+W^-$ and $h Z$ but their partial decay widths are rather suppressed compared to other channels, so that these channels cannot be used to test the model. For the sake of convenience, some formulae for the $S_1$ production cross sections are summarized in appendix\,\ref{app: S1 signals}.

\section{LHC direct search bounds on Heavy Fermion $X$}
\label{sec: bounds}

In this section, we consider a scenario where a $Z_2$-odd fermion $X$ with its hypercharge $Y_X$ decays into a stable and neutral $Z_2$-odd particle $\chi$, and discuss the current constraint on $X$ from 8\,TeV and 13\,TeV data at the LHC. We found in the previous section that the best fit value of $Y_X$ for the diphoton excess is $4/3$. For a $Z_2$-odd fermion with such an exotic hypercharge, we cannot write down any renormalizable interaction involving $X$, $\chi$ and SM particle(s). Interactions inducing $X$ decays are thus written by higher dimensional operators. For such an operator with a mass dimension as low as possible, ${\cal O}_F \sim (\bar{X} u^c)\,(\bar{\chi} u^c)/\Lambda^2$ (${\cal O}_S \sim (\bar{X} d^c)\,(\bar{u^c} d^c)\,\chi/\Lambda^3$) can be found for a fermionic (bosonic) $\chi$. Based on a naive dimensional analysis, the operators ${\cal O}_F$ and ${\cal O}_S$ lead to $\Gamma_X \sim (1/128 \pi^3)\,(m_X^5/\Lambda^4) \sim {\cal O}(1)$\,MeV and $\Gamma_X \sim (1/1024 \pi^5)\,(m_X^7/\Lambda^6) \sim {\cal O}(0.01)$\,MeV, respectively, when $\Lambda =1$\,TeV. The smallness of the $X$ width is required to enhance the diphoton signal as mentioned in the previous section and it is automatically guaranteed thanks to the exotic hypercharge.

As a result, the decay of $X$ proceeds as $X \to \chi + n$-jets, where $n \geq 2\,(3)$ when $\chi$ is a fermion (boson). We mainly consider the $n = 3$ case in this paper. On the other hand, we have to consider a vertex with the color structure of $\epsilon_{ijk} \bar{X}_i u^c_j u^c_k$ for the $n = 2$ case, but there are subtle points to simulate the color flow of the vertex. In order to estimate the efficiency of the signal detection in the $n = 2$ case, we thus have generated events using the decay $X \to q g \chi$. We have found that the result is similar to the one obtained in the $n = 3$ case.

The LHC signature of the pair production of $X$ is characterized by $\slashed{E}_T$ and jets:
\begin{eqnarray}
	pp \to X\bar{X} \to (\chi + jjj)\,(\chi + jjj) \qquad [\slashed{E}_T + {\rm jets}].
\end{eqnarray}
Conventional searches of such a process usually rely on a large missing momentum. When the heavy fermion $X$ is heavier enough than the stable particle $\chi$, a model with $m_X \sim 375$\,GeV is already disfavored by existing searches. The bound becomes weaker for a degenerate spectrum with $m_X \sim m_\chi$ as the decay products are too soft to be detected and the missing momentum tends to be small due to the back-to-back $\chi$ configuration. 

Mono-jet searches are then sensitive for such a degenerate mass spectrum, which utilizes the ISR-jet in the next leading order process. It provides a significant transverse momentum to the system of the undetectable particle $\chi$ (dark matter) pair as follows:
\begin{eqnarray}
	pp \to X\bar{X}\,j^{\rm ISR} \to (\chi + jjj)\,(\chi + jjj) + j^{\rm ISR}
	\quad
	[{\rm Large}~\slashed{E}_T + {\rm Hard~ISR~jet} + {\rm Soft~jets}].
\end{eqnarray}
We consider the collider limit on the $(m_X, \Delta m)$-plane, because $m_X$ mainly controls the $X\bar{X}$ pair production cross section through the strong interaction and $\Delta m \equiv m_X - m_\chi$ does the signal efficiency. Note that the mass difference $\Delta m$ also controls the thermal relic abundance of the dark matter particle $\chi$, as will be discussed in the next section. Mono-jet searches at 8\,TeV\,\cite{Aad:2014nra, Aad:2015zva} and $\slashed{E}_T +$ jets searches at 13\,TeV\,\cite{ATLAS-CONF-2015-062} reported by the ATLAS collaboration are particularly important to estimate the current bound on our scenario. Applying the above analyses to the simulated signal events, we draw several contours of their 95\%\,C.L. limits on the ($m_X$, $\Delta m$)-plane. We have used {\tt MadGraph5\_aMC@NLO}\,\cite{Alwall:2014hca} and {\tt Pythia6}\,\cite{Sjostrand:2006za} for the event generation. {\tt CheckMATE-1.2.2}\,\cite{Drees:2013wra} is used for the efficiency estimation, where {\tt Delphes3}\,\cite{deFavereau:2013fsa} and {\tt FastJet}\,\cite{Cacciari:2011ma, Cacciari:2008gp} are implemented for detector simulation and jet reconstruction, respectively. For the 13 TeV analysis, we follow the {\tt CheckMATE} convention and estimate the efficiency using the default {\tt Delphes3} detector card.

\begin{table}[t]
	\centering
	{\small
	\begin{tabular}{r|rrrrrrrrr}
	$m_X$\,[GeV] & 360 & 365 & 370 & 375 & 380 & 385 & 390& 395 & 400 \\
	\hline
	$\sigma$\,@\,\,8\,TeV\,[pb] & 4.35 & 4.01 & 3.70 & 3.41 & 3.15 & 2.91 & 2.69 & 2.49 & 2.31 \\
	$\sigma$@13\,TeV\,[pb] & 20.34 & 18.86 & 17.51 & 16.26 & 15.12 & 14.07 & 13.10 & 12.21 & 11.39 \\
	\hline
	\end{tabular}
	}
	\caption{\sl \small Pair production cross sections of the heavy fermion $X$ at NNLO for various $m_X$.}
	\label{tab: xs}
\end{table}

We have implemented a heavy colored fermion $X$ and a scalar dark matter $\chi$ using {\tt Feynrules}\,\cite{Alloul:2013bka} for the {\tt MadGraph} model file. Signal events are generated up to two additional jets and merged in the MLM-matching scheme\,\cite{Mangano:2002ea, Mangano:2006rw, Alwall:2007fs}. The NLO/NNLO production cross section of the $X$ pair is computed by {\tt Hathor-2.0}\,\cite{Aliev:2010zk} and used for the normalization of the events generated by {\tt MadGraph}. Numerical values used in our analysis are summarized in Table\,\ref{tab: xs} for various $X$ masses. The uncertainty of the cross section was estimated by changing the factorization scale, the renormalization scale and the parton distribution function, and it turns out to be less than 25\%\,(10\%) at NLO (NNLO)\,\cite{Aliev:2010zk}. To estimate the acceptance uncertainty, the theoretical error of the ISR distribution has to be considered. The error is often estimated as the deviation of the LO $X\bar{X}\,j$ matched cross section after the cut by changing the scale of the renormalization, factorization and emission vertex between 0.5 and 2 from the nominal value, keeping the normalization of the total cross section as the NLO one. We found the error of (the cross section) $\times$ (the acceptance) is dominated by the cross section uncertainty. It should be smaller if the NNLO fully differential cross section is available, however currently is not. We adopt 16\% theoretical error for (the cross section) $\times$ (the acceptance), which is the one quoted for the degenerate stop study at NLO\,\cite{Aad:2014nra}.

\begin{table}[t]
	\centering
	{\small
	\begin{tabular}{l|cccccc|rr}
	& $\slashed{E}_T$\,[GeV] & $p_{T\,j_1}$\,[GeV] & $\Delta\phi\,(j, \slashed{p}_T)$ & $n_j$
	& $\slashed{E}_T/\sqrt{H_T}$ & $m_{\rm eff}$\,(incl.) & $\sigma^{95\%}_{\rm obs}$ & Ref. \\
	\hline
	M2 & 340 & 340 & 0.4 & $\le 3$ & - & - & 28.4\,fb & \cite{Aad:2014nra} \\
	SR5 & 350 & 0.5$\slashed{E}_T$ & 1.0 & - & - & - & 21\,fb & \cite{Aad:2015zva} \\
	SR6 & 400 & 0.5$\slashed{E}_T$ & 1.0 & - & - & - & 12\,fb & \cite{Aad:2015zva} \\
	SR2jm & 200 & 300 & 0.4 & $\ge 2$ & 15\,GeV$^{1/2}$ & 1.2\,TeV & 21\,fb & \cite{ATLAS-CONF-2015-062} \\
	\hline
	\end{tabular}
	}
	\caption{\sl \small Signal regions and upper bounds on the signal cross sections at 95\%\,C.L. Here, $p_{T,\,j_1}$, $H_T$ and $m_{\rm eff}$\,(incl.) are the leading jet $p_T$, the scalar $p_T$ sum of all jets and $H_T + \slashed{E}_T$, respectively.}
	\label{tab: cuts}
\end{table}

Among various results obtained at the 8\,TeV LHC, the mono-jet searches\,\cite{Aad:2014nra, Aad:2015zva} are found to be particularly important for our scenario. In the former reference\,\cite{Aad:2014nra}, the following kinematical cuts have been adopted for all of the signal regions:
\begin{itemize}
	\setlength{\itemsep}{0cm}
	\item $\slashed{E}_T>150$\,GeV.
	\item At least a jet with $p_T > 150$\,GeV and $|\eta| < 2.8$.
	\item $n_j \le 3$ where $n_j$ is the number of jets with $p_T > 30$\,GeV and $|\eta|<2.8$.
	\item $\Delta\phi\,(j, \slashed{p}_T) > 0.4$ for each jet.
	\item No isolated leptons with $p_{T,\,\ell}>10$\,GeV.
\end{itemize}
Here, $\Delta\phi\,(j, \slashed{p}_T)$ is the azimuthal angle separation between the missing momentum and a jet selected by the cuts. Signal regions are then categorized into M1--M3 according to different cuts on $\slashed{E}_T$ and the leading jet $p_T$. On the other hand, in the latter reference\,\cite{Aad:2015zva}, the following kinematical cuts have been adopted for all of the signal regions: 
\begin{itemize}
	\setlength{\itemsep}{0cm}
	\item $\slashed{E}_T>150$\,GeV.
	\item $p_T/\slashed{E}_T > 0.5$ for the leading jet.
	\item $\Delta\phi\,(j, \slashed{p}_T) > 1.0$.
	\item No isolated leptons with $p_{T,\,\ell}>7$\,GeV.
\end{itemize}
The requirement on $\Delta\phi\,(j, \slashed{p}_T)$ is applied for each jet with $p_T > 30$\,GeV and $|\eta| < 4.5$. Signal regions are categorized into SR1--SR9 according to different cuts on $\slashed{E}_T$. Note that the signal regions are overlapped with each other, and thus statistically not independent. The upper limit on the maximum number of resolved jets and rather large $p_T$ values for the leading jet are required in the signal regions M1--M3. Those select only events which are close to pure mono-jet ones and thus gives a weaker bound compared to the one from SR5--SR6. Moreover, we have found that searches for squarks and gluinos using jets and missing momentum events at the 13 TeV LHC\,\cite{ATLAS-CONF-2015-062} have already set a severe constraint on our scenario. Following kinematical cuts have been adopted for all of the signal regions,
\begin{itemize}
	\setlength{\itemsep}{0cm}
	\item $\slashed{E}_T>200$\,GeV.
	\item $\Delta\phi\,(j, \slashed{p}_T) > 0.8$.
	\item No isolated leptons with $p_{T,\,\ell}>10$\,GeV.
\end{itemize}
We found that the most sensitive signal region for our scenario is SR2jm which are designed for the compressed spectra, where $n_j \ge 2$ for jets with $p_T > 50$\,GeV \& $|\eta|<2.8$ and the leading jet $p_T > 300$\,GeV are required to select mono-jet like events. We summarize in Table\,\ref{tab: cuts} the most sensitive signal regions in Refs.\,\cite{Aad:2014nra, Aad:2015zva, ATLAS-CONF-2015-062} for our scenario.

\begin{figure}[t]
	\centering
	\includegraphics[width=0.85\linewidth]{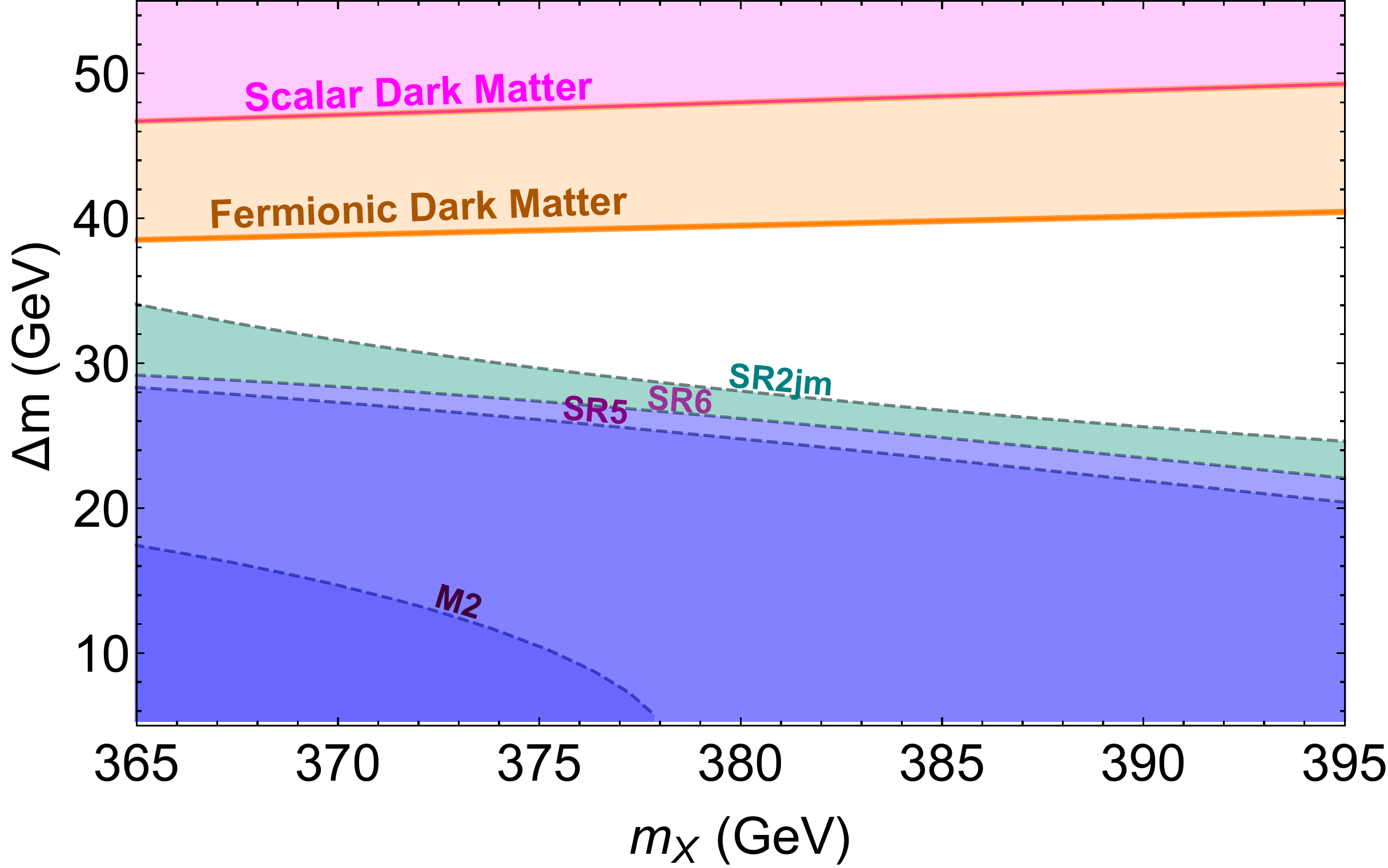}
	\caption{\sl \small Parameter regions excluded by the 8\,TeV LHC (M2, SR5 and SR6) and the 13\,TeV LHC (SR2jm) at 95\%\,C.L. on the $(m_X, \Delta m)$-plane. Systematic uncertainty of 16\% on the signal cross section is assumed. The region not favored from the dark matter ($\chi$) cosmological viewpoint is also shown for both the cases of scalar and the fermionic $\chi$. See text for more details.}
	\label{fig: exclusions}
\end{figure}

Our results are summarized in Fig.\,\ref{fig: exclusions}, where parameter regions excluded at 95\%\,C.L. are shown on the $(m_X, \Delta m)$-plane. Those are obtained using the prescription\,\cite{Read:2002hq}, $\sigma_{\rm sig} - 2\Delta \sigma_{\rm sig} < \sigma^{95\%}_{\rm obs}$, in each selected signal region, where $\sigma_{\rm sig}$ is the signal cross section after all the selection cuts applied in each signal region, while $\Delta \sigma_{\rm sig}$ is its error which is taken to be $16\%$, and $\sigma^{95\%}_{\rm obs}$ is the 95\%\,C.L. experimental upper limit on $\sigma_{\rm sig}$ quoted from the corresponding ATLAS analysis. Larger $\Delta m$ provides less mono-jet like events with $m_X$ being fixed, because additional jet activities reduce $\slashed{E}_T$ relative to the total activity of the events and thus reduce signal efficiencies. It can be seen that the parameter region with $\Delta m \gtrsim 30$\,GeV has not been excluded yet by the mono-jet searches when $m_X \sim 375$\,GeV. The signal regions SR5 and SR6 in Ref.\,\cite{Aad:2015zva} set much stringent constraints compared to the M2 region in Ref.\,\cite{Aad:2014nra}. The M2 region was indeed optimized for the scalar top search, as it requires strong criteria on the number of additional jets, leading to the reduction of multi-jet signal events in our case. This is because the mass scale of $X$ is higher than that expected in the original scalar top analysis, and the probability to have additional jets becomes higher in our case. The 13\,TeV limit from the SR2jm region is slightly stronger than the 8 TeV results. We find, however, the region with $\Delta m \gtrsim 30$\,GeV still survives for $m_X \sim 375$\,GeV.

In the same figure, we have also depicted the parameter region that is not favored from the cosmological viewpoint of $\chi$ (dark matter). When $\chi$ is a scalar\,(fermionic) particle, the region with $\Delta m \gtrsim$ 50\,(40)\,GeV is not favored, for the thermal relic density of $\chi$ exceeds the dark matter density observed today assuming that the self-annihilation cross section of $\chi$ is small. See the next section for more details. It can be seen that there is still an available region consistent with both the current LHC limits and the DM relic abundance. Interestingly, there is a 1$\sigma$ level excess observed in the SR2jm region against the estimated SM backgrounds and the systematic error is dominated. Our scenario is therefore compatible with the current situation and the entire parameter region can be probed once the systematic error is reduced by accumulating more events in near future.

\section{Cosmology of $\chi$}
\label{sec: cosmology}

We saw in the previous section that the mass difference between the dark matter particle $\chi$ and the new heavy fermion $X$ should be small enough but still larger than $\sim$30\,GeV to avoid the constraint from LHC direct searches. Such a mass difference is also favored from the viewpoint of dark matter cosmology, and this is the topic discussed in this section.

Since $\chi$ is degenerate with $X$ in mass, (co)annihilation processes among the two particles play crucial roles to determine the dark matter relic abundance at present universe. Let us first check whether or not the chemical equilibrium between the two particles is maintained during the freeze-out epoch. Since $X$ eventually decays into $\chi$ by emitting some SM particles, some interaction must exist between $X$ and $\chi$, where its reaction rate is parameterized by the decay width of $X$, namely $\Gamma_X$. There is an upper limit on $\Gamma_X$ so that it does not dominate the total decay width of the $X$-quarkonium, which reads $\Gamma_X \lesssim {\cal O}(1)$\,MeV. On the other hand, the width should be larger than the expansion rate of the universe during the freeze-out epoch, which is given by $\Gamma_X \gg H$ with $H \sim {\cal O}(10^{-16})$\,GeV being the Hubble constant during the epoch. These two conditions are thus easily satisfied simultaneously, and the two particles $\chi$ and $X$ can be assumed to be in the chemical equilibrium.

The relic abundance of the dark matter $\chi$ is then determined by so-called the thermally averaged effective annihilation cross section, which is in our scenario given as follows:
\begin{eqnarray}
\langle \sigma v \rangle =
	\sum_{ij} \langle \sigma_{ij} v \rangle \frac{g_i g_j}{g^2_{\rm eff}}
	(1 + \Delta_i)^{3/2} (1 + \Delta_j)^{3/2}
	\exp[-x\,(\Delta_i + \Delta_j)],
\end{eqnarray}
where $g_i$ is the spin and color degree of freedom for the particle `$i$', while $x = m_\chi/T$ and $\Delta_i = (m_i - m_\chi)/m_\chi$ with $m_\chi$, $T$ and $m_i$ being the dark matter mass, the temperature of the universe and the mass of the particle `$i$', respectively. The thermally averaged annihilation cross section between the particles '$i$' and `$j$' is denoted by $\langle \sigma_{ij} v \rangle$ with $v$ being the relative velocity between the two particles. The effective degree of freedom $g_{\rm eff}$ is defined by $g_{\rm eff}  = \sum_i g_i (1 + \Delta_i)^{3/2} \exp(-x \Delta_i)$. The index `$i$' runs among the dark matter $\chi$, the new vector-like quark $X$ and its anti-particle $\bar{X}$ in our setup. Annihilation cross sections $ \sigma_{\chi X}$ and $\sigma_{\chi \bar{X}}$ are negligibly small, because the interaction between the two particles are suppressed, as mentioned in the previous paragraph. The same reason is applied for cross sections $\sigma_{X X}$ and $\sigma_{\bar{X} \bar{X}}$. We also assume $\sigma_{\chi \chi} \ll \sigma_{X \bar{X}}$ in our analysis, which is justified in particular when $\chi$ is a fermion singlet under the SM gauge group, because all renormalizable interactions of such a particle are forbidden due to the SM gauge symmetry and the $Z_2$ symmetry to stabilize the dark matter particle. The effective annihilation cross section is therefore simply determined by the annihilation cross section between $X$ and $\bar{X}$, namely $\sigma_{X \bar{X}}$ in this setup.

Main contribution to the annihilation cross section $\sigma_{X \bar{X}}$ comes from QCD processes. Neglecting all the masses of SM particles, which is verified when $X$ is enough heavier than the SM particles, the effective annihilation cross section $\langle \sigma v \rangle$ eventually reads\footnote{We use in our numerical computation a more accurate formula for $\sigma_{X \bar{X}}$ including electroweak processes.}
\begin{eqnarray}
	\langle \sigma v \rangle \simeq
	2\,\frac{43 \pi \alpha_s^2}{27 m_X^2}
	\frac{36\,(1 + \Delta_X)^3 \exp(-2x \Delta_X)}
	{\left[g_\chi + 12\,(1 + \Delta_X)^{3/2} \exp(-x \Delta_X)\right]^2}.
\end{eqnarray}
According to the method developed in Ref.\,\cite{Griest:1990kh} and using the cross section $\langle \sigma v \rangle$ computed, the relic abundance of the dark matter at present universe is obtained by solving the Boltzmann equation. Our result is shown in Fig\,\ref{fig: exclusions} as a thin orange (pink) band and a shaded region with the same color, where the dark matter particle is assumed to be a fermion (scalar). When $m_X$ and $\Delta m \equiv m_X - m_\chi$ are inside the band, the correct relic abundance observed today, $\Omega_{\rm DM} h^2 \simeq  0.112$, is reproduced at 95\% confidence level neglecting systematic errors associated with theory predictions. The relic abundance exceeds the observed value in the shaded region, so that it is excluded. As can be seen in the figure, the mass difference $\Delta m$ of about 40\,(50)\,GeV is favored for a fermionic (scalar) dark matter $\chi$, which is within the region evading the direct heavy fermion $X$ searches at the LHC.

\section{Outlook}
\label{sec: outlook}

We have shown that the non-relativistic bound state of a pair of the colored particles $X$ with $m_X\sim 375$\,GeV and $Y_X=4/3$ can be responsible for the 750\,GeV diphoton excess. This scenario is consistent with current LHC data from $p p \to X \bar{X}$ if $X$ decays into a stable neutral particle $\chi$ and multiple soft jets though higher dimensional operators. The mass difference between the new particles $\Delta m = m_X - m_\chi$ must be small to avoid hard jets + missing $E_T$ constraints, but must also be large enough to avoid mono-jet search constraints. $\Delta m > 30$\,GeV is required for $m_X \sim 375$\,GeV. On the other hand, the stable particle $\chi$ in the scenario can be a dark matter in our universe. Even assuming the pair annihilation of $\chi$ is small, its thermal relic density can be small enough if $\Delta m < 40\,(50)$\,GeV when $\chi$ is a fermion (scalar). Collider constraints on mono-jet searches would be more stringent in future, and it will excludes or proves the entire region satisfying this cosmological constraint.

The diphoton signature is currently not significant enough statistically. Even if the diphoton excess does not survive in future, our work is still useful to constrain a class of scenario where a dark matter couples to a heavy colored fermion but does not couple to the SM sector at leading order. For integrated luminosity of 3000 fb$^{-1}$, which is in the scope of so-called the high luminosity LHC, the LHC can give a factor of 1/30 more stringent limit on the signal cross section $\sigma( p p \to S_0 \to \gamma \gamma)$ than the current one in purely statistical consideration. This is sufficient to access another type of the heavy colored fermion $X'$ with $Y_{X'} = 2/3$, which has the same charge of the standard model top quark. The search for the mono-jet signal should also constrain the scenario, however its constraint may suffer more from theoretical and systematical errors compared with the case of the diphoton resonance search. 

\vskip 0.5cm
\noindent
{\bf Acknowledgments}\\[0.1cm]
\noindent
This work is supported by the Grant-in-Aid for Scientific research from the Ministry of Education, Science, Sports, and Culture (MEXT), Japan No. 26104009 (S.M.), No. 23104006 (M.M.N) and No. 26287039 (S.M. and M.M.N.), as well as by the World Premier International Research Center Initiative (WPI), MEXT, Japan. The work of K.I. is supported in part by a JSPS Research Fellowships for Young Scientists.

\appendix

\section{Fitting functions for $|\psi_0(0)|$ and $E_0$}
\label{app: wave function}

The wave function at origin, $|\psi_0(0)|$, and the corresponding energy eigenvalue $E_0$ of the bound state $S_0$ are required to compute the signal cross section for the diphoton excess. Fitting functions of these two quantities for various hypercharges $Y_X$ are given by
\begin{eqnarray}
	|\psi_0 (0)|  = \sum_{n = 0}^4\,a_n\,[\,\ln\,(m_{S_{0}}/750\,{\rm GeV})\,]^n,
	\quad\quad
	E_0 = \sum_{n = 0}^4\,b_n\,[\,\ln\,(m_{S_{0}}/750\,{\rm GeV})\,]^n,
	\label{eq: fittings}
\end{eqnarray}
where the coefficients $a_n$ and $b_n$ for various $Y_X$ are given in Table\,\ref{tab: fittings}. It is worth emphasizing here that our fitting results with $Y_X = 0$ are consistent with those in Ref\,\cite{Hagiwara:1990sq}. In fact the difference is at most four percent in the range of 200\,GeV $\leq m_{S_{0}} \leq$ 1\,TeV. Note also that our fitting functions were verified to work up to the case with 200\,GeV $\leq m_{S_{0}} \leq$ 2\,TeV.

\begin{table}[t]
	\centering
	{\small
	\begin{tabular}{c|ccccc|ccccc}
	$Y_X$ & $a_0$ & $a_1$ & $a_2$ & $a_3$ & $a_4$ & $b_0$ & $b_1$ & $b_2$ & $b_3$ & $b_4$ \\
	\hline
	0 & 87.78 & 114.4 & 76.85 & 37.76 & 10.71 & 4.119 & 2.458 & 0.9314 & 0.2429 & 0.04078 \\
	1/3 & 88.44 & 115.3 & 77.54 & 38.14 & 10.82 & 4.145 & 2.481 & 0.9416 & 0.2461 & 0.04143 \\
	2/3 & 90.44 & 118.1 & 79.64 & 39.30 & 11.17 & 4.226 & 2.552 & 0.9726 & 0.2557 & 0.04341 \\
	1 & 93.82 & 122.9 & 83.20 & 41.25 & 11.76 & 4.363 & 2.672 & 1.026 & 0.2721 & 0.04682 \\
	4/3 & 98.64 & 129.8 & 88.28 & 44.05 & 12.61 & 4.559 & 2.845 & 1.102 & 0.2960 & 0.05180 \\
	5/3 & 105.0 & 138.8 & 95.01 & 47.76 & 13.73 & 4.822 & 3.077 & 1.206 & 0.3285 & 0.05858 \\
	2 & 112.6 & 150.7 & 104.5 & 51.83 & 14.40 & 5.162 & 3.366 & 1.322 & 0.3812 & 0.07999 \\
	\hline
	\end{tabular}
	}
	\caption{\sl \small Coefficients $a_n$ and $b_n$ for the fitting functions for $\phi_0(0)$ and $E_0$ in Eq.\,(\ref{eq: fittings}).}
	\label{tab: fittings}
\end{table}

\section{Signals from $S_1$ bound state}
\label{app: S1 signals}

The spin one bound state $S_1$ is produced dominantly by quark-antiquark collisions at the LHC, and then decays into various pairs of SM particles, such as a lepton pair. As in the $S_0$ case, the signal cross section of the $S_1$ bound state into the $x\bar{x}$ final state is given by
\begin{eqnarray}
	\sigma (pp \to S_1 \to x\bar{x}) = 
	\frac{3\,K_1}{s\,m_{S_1}}
	\sum_q
	\frac{\Gamma^{(1)}_{x\bar{x}}\,\Gamma^{(1)}_{q\bar{q}}}{\Gamma^{(1)}_{\rm tot}} 
	\left[ \frac{4\pi^2}{9} \int dx_1 dx_2 \delta(x_1 x_2 - m_{S_1}^2/s)\,f_q(x_1) f_{\bar{q}}(x_2) \right],
\end{eqnarray}
where $m_{S_1}$ is the mass of the bound-state. Since $S_1$ exactly degenerates $S_0$ in mass at leading order, it is the same as $m_{S_0}$. The quark and antiquark PDFs inside a proton are denoted by $f_q(x)$ and $f_{\bar{q}}(x)$. According to MSTW2008NLO\,\cite{Martin:2009iq}, the parenthesis of the right hand side gives, e.g. 158, 89 and 7.2 for $q = u$, $d$ and $s$, respectively, at $\sqrt{s} = 8$\,TeV with $m_{S_1}$ being 750\,GeV\,\cite{Franceschini:2015kwy}. The so-called $K$-factor is denoted by $K_1$ in the above cross section. 

The total decay width of the $S_1$ bound state and its partial decay width into the $x\bar{x}$ final state are denoted by $\Gamma^{(1)}_{\rm tot}$ and $\Gamma^{(1)}_{x\bar{x}}$, respectively, and their explicit forms are given by
\begin{eqnarray}
	\Gamma^{(1)}_{\rm tot} = 82 \frac{\pi\,Y_X^2\,\alpha^2\,|\psi_1(0)|^2}{c_W^4\,m_{S_1}^2},
	\qquad
	\Gamma^{(1)}_{x\bar{x}} = c_{x\bar{x}} \frac{\pi\,Y_X^2\,\alpha^2\,|\psi_1(0)|^2}{c_W^4\,m_{S_1}^2}.
\end{eqnarray}
The wave function $\psi_1(0)$ is equal to $\psi_0(0)$, because the Schr\"odinger equation for $S_1$ is exactly the same as the one for $S_0$ at leading order. The coefficient $c_{x\bar{x}}$ are 20, 10, 34/3 and 10/3 when $x\bar{x} = \ell^+\ell^-$, $\tau^+\tau^-$, $u\bar{u}$\,($= c\bar{c} = t\bar{t}$), $d\bar{d}$\,($= s\bar{s} = b\bar{b}$), respectively.

\bibliographystyle{aps}
\bibliography{refs}

\end{document}